\documentclass[sn-mathphys]{sn-jnl}% Math and Physical Sciences Reference Style
\usepackage{hyperref}
\usepackage{url}
\jyear{2021}%

\theoremstyle{thmstyleone}%
\theoremstyle{thmstyletwo}%

\theoremstyle{thmstylethree}%

\begin{document}

\title[Automatic Single Cough Segmentations]{Comparing Hysteresis Comparator and RMS Threshold Methods for Automatic Single Cough Segmentations}


\author*[1]{\fnm{Bagus Tris} \sur{Atmaja}}\email{b-atmaja@aist.go.jp}

\author[2]{\fnm{Zanjabila} \sur{}}\email{zanjabilaabil@gmail.com}

\author[2]{\fnm{Suyanto} \sur{}}\email{suyanto@ep.its.ac.id}

\author[1]{\fnm{Akira Sasou} \sur{}}\email{a-sasou@aist.go.jp}

\affil[1]{\orgname{National Institute of Advanced Industrial Science and Technology}, \city{Tsukuba}, \postcode{3058560}, \country{Japan}}

\affil*[2]{\orgdiv{Department of Engineering Physics}, \orgname{Sepuluh Nopember Institute of Technology}, \orgaddress{\street{ITS Sukolilo Campus}, \city{Surabaya}, \postcode{60111}, \state{Jawa Timur}, \country{Indonesia}}}

% \affil[3]{\orgdiv{Department}, \orgname{Organization}, \orgaddress{\street{Street}, \city{City}, \postcode{610101}, \state{State}, \country{Country}}}


\abstract{Research on diagnosing diseases based on voice signals is rapidly increasing, including cough-related diseases. When training the cough sound signals into deep learning models, it is necessary to have a standard input by segmenting several cough signals into individual cough signals. Segmenting coughs could also be used to monitor trends of cough-related disease by counting the number of coughs. Previous research has been developed to segment cough signals from non-cough signals. This research evaluates the segmentation methods of several cough signals from a single audio file into several audio files containing a single file. We evaluate three different methods, including manual segmentation as a baseline and two automatic segmentation methods: hysteresis comparator and root mean square (RMS) methods. The results by two automatic segmentation methods obtained precisions of 73\% (hysteresis) and 70\% (RMS) compared to 49\% by manual segmentation. The agreements of listening tests to count the number of correct single-cough segmentations show fair and moderate correlations for automatic segmentation methods and are comparable with manual segmentation.}




\maketitle

\section{Introduction}
% Voice-based disease diagnosis is a rapidly increasing field.
Utilizing acoustic signals as a tool for health diagnosis has been started with voice-related diseases (pathological speech). Arifianto et al. \cite{Arifianto2002a} analyzing pathological speech through time-frequency analysis. The method is applied to detect vibrational abnormalities of the vocal folds for larynx diseases. They utilized a similar method to classify the degree of severity of speech disorders via a time-varying autoregressive method \cite{Arifianto2004}. The growth of speech pathology from acoustic analysis has gained more interest through the rapid development of machine learning and signal processing (ML-SP) methods \cite{Gupta2016}. The growth of ML-SP opens the possibility of using speech signals as a tool for other diagnoses not limited by pathological speech. The most noticeable case currently is the use of speech signal analysis (instead of cough) to predict the presence of Corona Virus Desease 2019 (COVID-19) signature in the vowel sounds \cite{Bartl-Pokorny2021a,Vahedian-azimi2021,Shimon2021}. Not only for physical health diagnosis (e.g., dysphonia detection \cite{Lad2019}, lung sound analysis \cite{Singh2023}), but the use of voice analysis is also useful for mental health diagnosis (e.g., monitoring of emotions \cite{Atmaja2022g,Atmaja2022h,Jayanthi2022,Selvan2023}).

Cough is a common symptom of several diseases, particularly respiratory diseases. Acoustic analysis has several advantages for cough analysis as the main phenomenon of cough is the explosive sound (which is clearer than regular speech). By using acoustics, it can be predicted whether it is dry or wet cough \cite{Swarnkar2013}, there is asthma from cough sounds \cite{khassaweneh2013},  whether it contains pertussis or not \cite{Pramono2016}. Recent advancements in deep learning methods offer a simpler way to train a large corpus of cough sounds for such a disease, e.g., COVID-19 detection through cough sounds \cite{Chaudhari2020,DariciHaritaoglu,Islam2022,Orlandic2021,Hamidi2023}.

% cough segmentation
It may be necessary to have standard inputs by segmenting several cough sounds in a single audio file into individual cough sounds, such as training cough sounds in a deep learning method. Machine and deep learning methods have been found to be sensitive to the variations of the input data; hence, such pre-processing is necessary. The process includes standardization, normalization, and similar. Previous research has shown that these pre-processing techniques, including on-the-fly processing, improve and accelerate the performance of deep learning methods (e.g., batch processing \cite{Ioffe2015}). However, the previous pre-processing techniques only focused on normalization and standardization. To the best of our knowledge, no effort has been found to standardize several coughs into individual coughs, which will benefit the training of cough sounds for deep learning methods.

Aside from the normalization and standardization of sound signals for deep learning methods, the research on cough segmentation has been growing with the aim of extracting cough signals from non-cough signals (noises such as music, speech, and other sounds). In \cite{Amrulloh2015}, the authors proposed to extract cough regions by mathematical features such as non-Gaussianity, Shannon entropy, and cepstral coefficients. They achieved 97.3\% accuracy for cough sounds recorded in pediatric wards. While this method is not intended to extract individual coughs, a theoretical framework for early COVID-19 detection through cough sounds is proposed in \cite{Belkacem2021} along with cough detection and segmentations. However, there is no clear explanation for this theoretical idea (without results) nor steps to segment coughs into individual coughs. The authors only share the clue that the segmentation phase will be performed based on pre-defined thresholds.

% single-cough segmentation
In need of an empirical study for segmenting cough sounds into individual or single coughs (in this paper, the terms single coughs and individual coughs are used interchangeably), we evaluated several methods of cough segmentations in this study. This individual segmentation is important for further analysis, particularly when the next step is to train a deep-learning model. Furthermore, we believe that other cough-related analyses (e.g., classification of cough type, diagnosis) will benefit from these individual cough segmentations and will perform better if it is conducted based on the segmentation results.

Segmenting coughs into individual coughs is important because it can help healthcare professionals more accurately diagnose a person's illness or condition. By analyzing individual coughs separately, doctors and other medical professionals can better understand the characteristics of the cough and how it has changed over time. This can provide important information about a person's health, such as whether they have a respiratory infection or other medical condition. Additionally, segmenting coughs into individual coughs can help healthcare professionals monitor a person's progress and determine the effectiveness of any treatments that have been prescribed. Overall, segmenting individual coughs can provide valuable information for diagnosing and treating a person's illness or condition.

Cough segmentation will be useful in the future for calculating the number of coughs (Cough counting/cough frequency) to help doctors determine the disease.So far, the best method for recognizing cough and calculating cough frequency is unclear and most are done manually by humans using the ear \cite{Turner2014}. According to Jocelin et al. \cite{Hall2020}, cough frequency is now the gold standard for trying new treatments for chronic cough, investigating areas of infection in tuberculosis, and as a marker of chronic obstructive pulmonary disease (COPD) recovery. In addition, according to Arietta et al. \cite{Spinou2014}, monitoring the number of cough frequencies is also useful for assessing the results of therapy. Therefore, the cough segmentation method is important in cough calculation and monitoring in the future, especially for monitoring infectious respiratory diseases. An automation system is needed to calculate the number of cough frequencies.

There are several methods that can be used to segment cough sounds. Some of the most common methods include using machine learning algorithms to identify and classify cough sounds, analyzing the frequency and intensity of the sound, and comparing the cough sound to a reference database of known cough sounds. Additionally, some healthcare professionals may use visual analysis techniques to manually segment cough sounds, such as by listening to the sounds and marking the beginning and end of each cough. Ultimately, the specific method used to segment coughs will depend on the availability of technology and the preferences of the healthcare professionals involved. This paper provides an overview of the different methods that can be used to segment cough sounds and discusses the advantages and disadvantages of each method.

In a concrete way, this paper contributes to the evaluation of three segmentation methods for extracting individual cough regions. The first method is manual segmentation by detecting the silent regions. The second method is an automatic segmentation by detecting cough regions within two thresholds, lower and upper, with a hysteresis comparator. The third method is an automatic segmentation that employs a pre-defined single threshold to compare with the root mean square (RMS) of frames of the signals. The evaluations are measured by two metrics. Fleiss' Kappa measures inter-evaluator agreement for labeling individual cough and non-individual cough. The precision measures the percentage of correct single-cough segmentations over the total outputs of segmentation methods.

\section{Problem statement}
\begin{figure}
  \centering
  \includegraphics[width=0.87\textwidth]{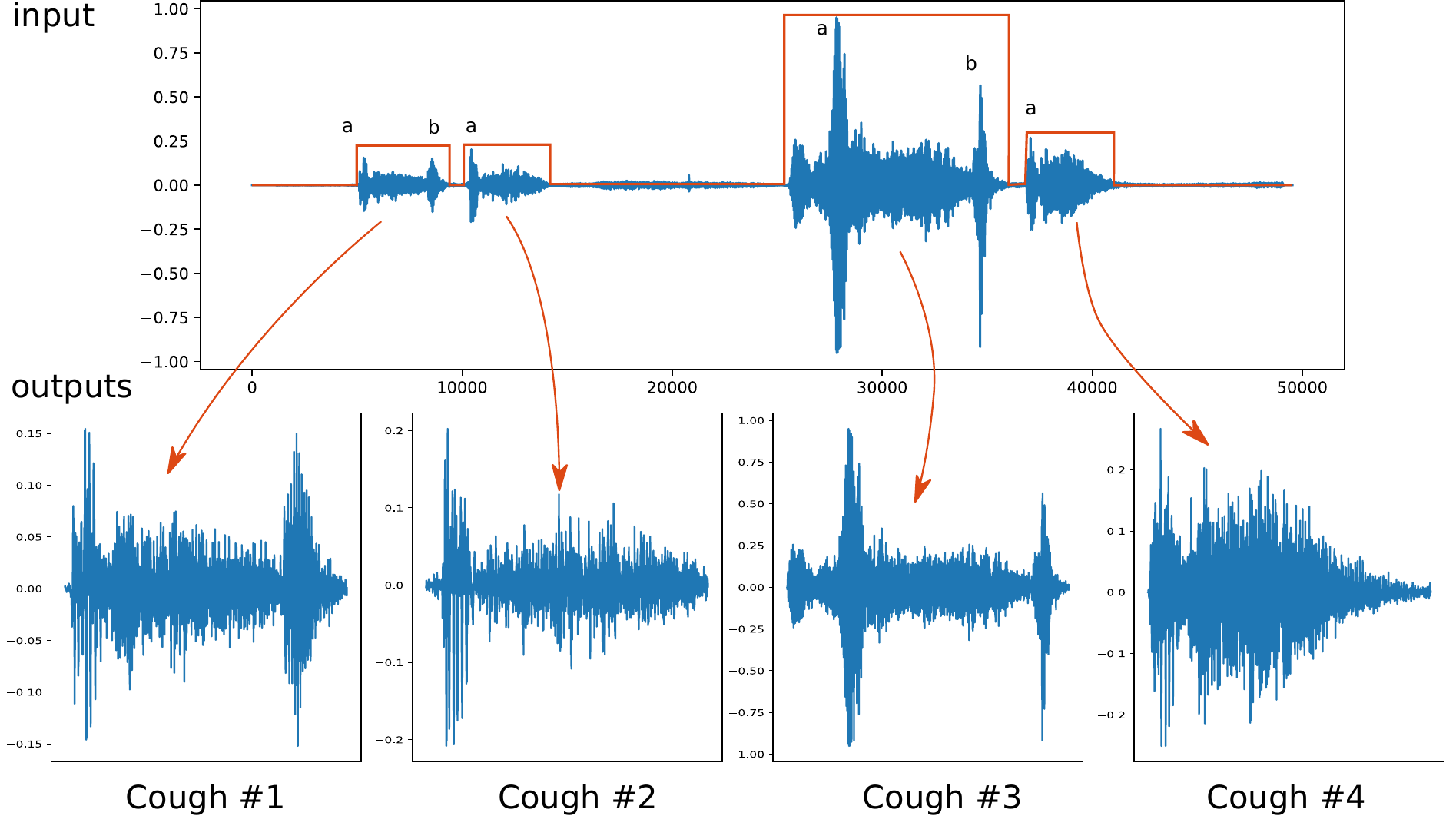}
  \caption{Segmentation of a raw cough sound containing several coughs (top) into individual coughs (bottom)}
  \label{fig:segmentation}
\end{figure}

The main problem of this study is to segment a single audio file containing several coughs into several audio files containing an individual cough for each file  (exported as WAVE files). Figure \ref{fig:segmentation} illustrates the segmentation process for both manual and automatic segmentation. In this case, the raw input is a single audio file with four coughs. The perfect segmentation method will produce four individual coughs. The input audio file and its annotation into four coughs, for that example, is obtained from the reference \cite{Turner2014}.

Notice that in Figure \ref{fig:segmentation} we used two phases to count a cough cycle: the initial explosive phase (a) and the final voiced phase (b). This counting is based on the study on how to count cough \cite{Turner2014}, where the listening tests to label segmentation results in this study are based on. Other references \cite{Lee2017, Hall2020} suggested three phases for one cycle of cough (explosive, intermediate, and voiced). In the case of two phases, the explosive and intermediate phases are combined into one explosive phase. While the first explosive phase is mandatory, the intermediate and voiced phases are not always present. Hence, as shown in Figure \ref{fig:segmentation}, the segmentation process for the example cough sound resulted in four individual coughs containing two coughs with phase \verb|a| and \verb|b|, and two coughs with phase \verb|a| only.

\section{Methods}
\begin{figure*}[htbp]
  \centering
  \includegraphics[width=.95\textwidth]{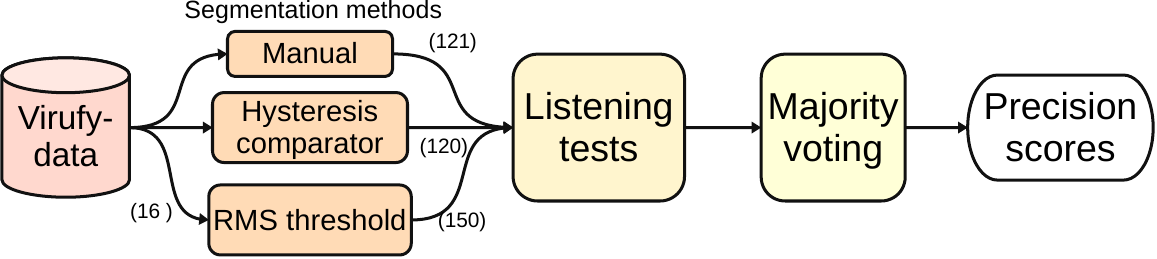}
  \caption{Flow of research from the raw data (virufy-data) to the results (precision scores). The numbers inside the bracket indicate the number of audio files from unsegmented (left) to segmented (right).}
  \label{fig:method}
\end{figure*}

We employed a research methodology as shown in Figure \ref{fig:method}. The main blocks of research are the dataset, the segmentation methods, subjective evaluations, and evaluation metrics. The processes are performed in order. First, we segmented the dataset (virufy-data) with the hysteresis comparator and RMS threshold methods. The (manual) segmented files from the virufy-data act as a baseline. The output of segmentations (including the manual/baseline) is evaluated by annotators through the listening test as subjective evaluations. The ratings from five annotators are aggregated with the majority voting to obtain a single label (single cough or non-single cough). Finally, the precision score of each segmentation method can be calculated by counting the number of single-cough and non-single-cough signals. Each of these blocks is described below.

\subsection{Dataset}
We utilized the virufy-data dataset \cite{Chaudhari2020} available at https://github.com/virufy/virufy-data. This dataset is intended to ``develop an AI-based method that accurately predicts COVID-19 infection" from crowdsourced cough audio samples recorded on smartphones \cite{Chaudhari2020}. The original dataset contains cough samples from 16 patients in mp3 format with a 48000 Hz of sampling rate in a single channel (mono). The audio files are already labeled with \textbf{pos}itive (``pos") and \textbf{neg}ative (``neg") labels for COVID-19 evaluation purposes. We do not use positive and negative labels in this research. Instead, we performed subjective evaluations for obtaining true labels of single-cough segments for manual segmentation provided by virufy-data dataset and our proposed automatic segmentation methods. The original clinical dataset contains 16 audio files (one audio file per patient).

\subsection{Segmentation methods}
In this subsection, we introduce three segmentation methods evaluated in this study: manual segmentation, hysteresis comparator, and RMS threshold. 

\subsubsection{Manual segmentation}
The (presumed) manual segmentations are provided in the original dataset \cite{Chaudhari2020}. The authors have provided 121 segmented cough samples from 16 patients (one audio file per patient). The segmented audio files were created by identifying periods of relative silence in the recordings and separating cough regions based on those silences. The audio segments that were not coughs or had a large number of background noise are removed \cite{virufy-data}. By this manual segmentation method, we assumed that the segmented audio files were intended to extract individual cough regions. The number of segmented audio files by this manual segmentation is 121 files.

\subsubsection{Hysteresis comparator}
We evaluated two automatic single-cough segmentation methods. First, we evaluated cough segmentation based on methods provided in the previous cough detection research \cite{Orlandic2021}. In this hysteresis comparator, we set low and high threshold multipliers as default, i.e., 0.1 and 2.0, to multiply with the RMS value, which is calculated from overall samples in the given audio signal. The hysteresis comparator identified cough regions of the signal with rapid spikes in power \cite{Orlandic2021}. We set the minimum cough length to 200 ms according to the reference \cite{Lee2017} and the padding to 0 ms to cut any signals not related to cough regions. The number of segmented individual coughs by this hysteresis comparator method is 120 files.

\subsubsection{RMS threshold}
In this automatic segmentation method with root mean square (RMS) threshold, we performed a simple segmentation to detect single cough regions from signals based on a specific threshold from the RMS value. We set the threshold to 0.09 from the normalized RMS energy. First, we divided the input audio files into frames of 42 ms (2048 samples of 48000 Hz sampling rate). If a frame in the evaluated signal is higher than the threshold, we count the samples inside that frame as cough regions and counting. The counter will stop if the frame is below the threshold. We limited the minimum length to 300 ms and the maximum length to 3 s based on empirical observations. We added three samples before and after cough signals to extend the original cough regions. The number of segmented individual cough files by this RMS threshold is 150 files.

\subsection{Subjective evaluation}
We asked six undergraduate students of Engineering Physics ITS to evaluate the segmented audio files produced by manual and automatic transcriptions through listening tests and displayed waveforms. First, the evaluators/annotators are asked to read the reference paper about how to count cough \cite{Turner2014}. Second, we provided them with the example of an audio file containing four coughs that emphasized detecting single-cough signals. This stage requires annotators to listen and watch the displayed waveform. Third, the evaluators are asked to judge whether the given audio file contains a single cough (labeled with "1") or a non-single cough (labeled with ``0"). One of the annotations was removed due to high variability with other annotators, resulting in five evaluated raters. A majority voting method from five annotators chose the most frequent labels for calculating the performance of the segmentation methods.

\subsection{Metrics}
We employed two metrics to evaluate our evaluation of single-cough segmentation methods. The first metric is the Fleiss' Kappa to measure the agreement of listening tests to count the number of correct single-cough segmentations. The second metric is the precision of the single-cough segmentation. Note that we compared the Fleiss' Kappa scores across different algorithms since the number of the segmented files for each algorithm differs from each other. 

\subsubsection{Fleiss' Kappa}
Fleiss' Kappa \cite{Fleiss1971} is utilized to measure inter-evaluator agreement since the evaluators annotate two categories (rather than counting the number of coughs), whether the listened and displayed audio file contains a single cough or not (two coughs or more, noises, or others). This Fleiss' Kappa is an extension of Cohen's Kappa, which is only able to measure the agreement for category items from two raters. It is also similar to the interclass correlation coefficient (ICC) agreement that works for measuring agreement for the annotation of continuous numbers.

% // fleiss formula and Interpretation
Let $P$ be the average of individual agreement per sample and $P_e$ be the average sum of the squared proportion of all assignments/evaluations. Fleiss' Kappa is formulated as,
\begin{equation}
  \kappa=\frac{\bar{P}-\bar{P_e}}{1-\bar{P_e}},
\end{equation}
\noindent where,
\begin{align}
\bar{P} &= \frac{1}{Nn(n-1)} \biggl(\sum_{i=1}^{N} \sum_{j=1}^{k} n_{ij}^2 - Nn \biggr), \\
\bar{P_e} &= \frac{1}{N} \sum_{j=1}^{k} p_j^2,\\
p_j &= \frac{1}{Nn} \sum_{i=1}^{N} n_{ij}, \\
\sum_{j=1}^{k} p_j &= 1.
\end{align}
The $p_j$ in eq. (4) is the proportion of all assignments, $N$ in eqs. (2), (3), and (4) is the total number of samples (segmented files) with index $i$, and $n$ in eqs. (2) and (4) is the number of annotators with index $j$. The range of $\kappa$ is on a scale of 0-1, with 0 for perfect disagreement and 1 for perfect agreement. To interpret the $\kappa$ value, we follow the convention for Cohen's kappa \cite{Landis1977} in Table \ref{tab:interpret}.

\begin{table}[htbp]
  \caption{Interpration of Fleiss' Kappa based on \cite{Landis1977}}
  \centering\begin{tabular}{c c}
  \hline
    $\kappa$ &	Interpretation \\
    \hline
    ~~~~$<$ 0 	        & Poor agreement \\
    0.01 – 0.20 	& Slight agreement \\
    0.21 – 0.40 	& Fair agreement \\
    0.41 – 0.60 	& Moderate agreement \\
    0.61 – 0.80 	& Substantial agreement \\
    0.81 – 1.00 	& Almost perfect agreement \\
  \hline
  \end{tabular}
  \label{tab:interpret}
\end{table}
% accuracy formula, equal to tpr and sensitivity

\subsubsection{Precision}
We reported precision or positive predictive value to evaluate the performance of cough segmentation methods. This metric is equal to the accuracy for the single-cough category since the output of the segmentation methods should contain only single-cough segments. Aside from single-cough predictions as true positive (TP), non-single-cough signals (e.g., several coughs, noises, or speeches) are observed as false positive (FP) after subjective test evaluations of segmented files. The precision is formulated as
\begin{equation}
  Precision = \frac{TP ~(single-cough)}{ TP + FP ~(non-single-cough)}.
\end{equation}
The labels of TP and FP came from the majority voting of five annotators in the previous subjective evaluation. Precision is the main metric to evaluate the performance of segmentation methods in this study.

\subsection{Open repositories}
All research methods, including the segmentation methods (python), listening test results (spreadsheets), and calculation of metrics (gnumeric) are open to the public in the following repository: https://github.com/bagustris/detect-segment-cough. The original and segmented audio files (in format MP3 and WAV, for demo) are available in the following repository: https://github.com/bagustris/virufy-data.

\section{Results and Discussion}
We divide our results into three subsections: annotator agreement, the precision of individual cough segmentations, and the error analysis of the cough automatic cough segmentation methods (compared to manual segmentation). The first subsection evaluates the reliability of the listening test to obtain the true labels for the latter subsection. The latter shows the performance of evaluated individual cough segmentations. The last subsection shows in which cases the automatic cough segmentation methods (RMS, hysteresis, or both) failed to segment the cough sounds.

\subsection{Annotator agreement}
We present our results in the scores of annotators' agreement (Table \ref{tab:kappa}) based on Fleiss' Kappa using eq. (1) and precision of individual cough segmentations (Table \ref{tab:acc}). The first table evaluates the reliability of the listening test to obtain the true labels for the latter table. The latter shows the performance of evaluated individual cough segmentations.

Table \ref{tab:kappa} shows that the agreement between annotators varies among three different segmentation methods. Since the number of resulting segmented files also varies between segmentation methods (i.e., 120 vs. 121 vs. 150, see Table \ref{tab:acc}), it is hard to compare directly between each method. The closest comparisons suggest that a simple segmentation with an RMS threshold obtained a better agreement among annotators. RMS threshold resulted in the largest number of segmented files, which may lead to a higher agreement since the probability of more individual coughs is greater than a smaller number of segmented files. While it is possible to limit the number of segmented files (output of segmentation) for the same comparison condition, one segmentation algorithm may perform optimally under this limitation. One may suggest another metric to evaluate the agreement among annotators in the future.

\begin{table}[htbp]
  \caption{Fleiss' Kappa rating of listening tests for two categories (single-cough vs. others) from five evaluators}
  \centering\begin{tabular}{l c c}
  \hline
  Segmentation method & Fleiss' Kappa & Interpretation \\
  \hline
  Manual segmentation & 0.345 & fair \\
  Hysteresis comparator & 0.246 &  fair \\
  RMS threshold & 0.486 & moderate \\
  \hline
  \end{tabular}
  \label{tab:kappa}
\end{table}

\subsection{Precision scores}
Table \ref{tab:acc} shows the performance of evaluated individual cough segmentation methods in terms of precision score. It can be inferred clearly that evaluated automatic segmentation methods outperform the manual segmentation method with silence criteria. Both hysteresis comparator and RMS threshold segmentation methods achieve comparable precision scores despite the different number of segmented files (150 vs. 120). Using two thresholds, as in the hysteresis comparator, seems to achieve a better performance since it limits individual cough based on two conditions (upper and lower thresholds). The number of segmented coughs between the manual segmentation and hysteresis comparator is also similar (121 vs. 120), highlighting the similarity between the two segmentation methods. The actual number of individual coughs among segmentation methods should be similar since it is a count of individual coughs within the same input audio file. Nevertheless, the number of segmented files by the RMS threshold is larger than others, showing a different mechanism of segmenting cough regions by this method.

\begin{table}
  \caption{Precision of single-cough segmentation methods over all segmented data; the used labels are from majority voting of five evaluators; N = number of segmented files.}
  \centering\begin{tabular}{l c c c c}
  \hline
  Segmentation method & N & Single-cough & Precision \\
  \hline
  Manual segmentation & 121 & 60 & 49.59\% \\
  Hysteresis comparator & 120 & 88 & 73.33\% \\
  RMS threshold & 150 & 105 & 70.00\% \\
  \hline
  \end{tabular}
  \label{tab:acc}
\end{table}

These findings on precision and agreement scores of evaluated segmentation methods open up new directions in cough-related research. The first is on the development of high-precision and high-agreement individual cough segmentation methods. The second is on the potential of automatic individual cough counter by employing individual cough segmentation methods. A non-automatic cough counter is still the standard gold method of quantifying the number of individual coughs \cite{Turner2014}.

\subsection{Error analysis}
\begin{figure}
  \centering
  \includegraphics[width=0.87\textwidth]{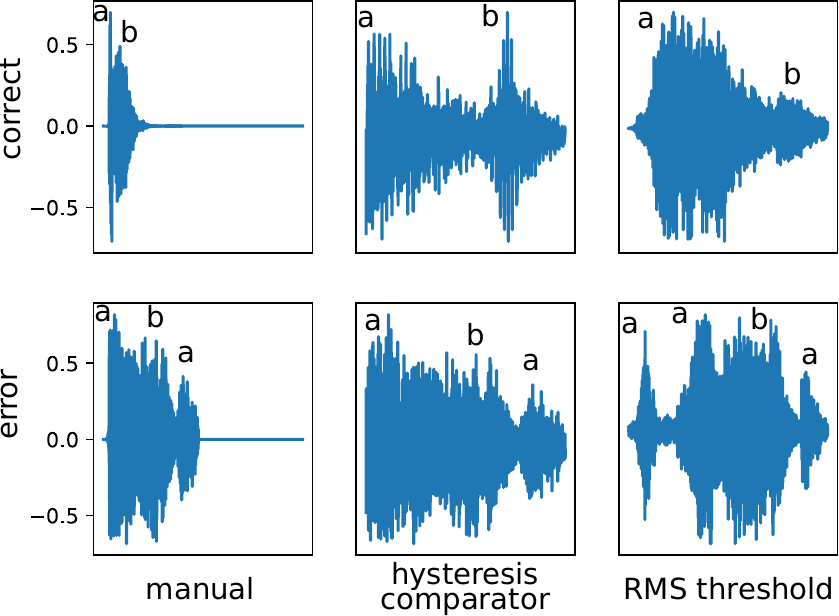}
  \caption{Example of correct (top) and error (bottom) of individual segmentation methods: in error cases, all segmentation methods result in more than one segmented cough; a: explosive phase, b: voiced phase}
  \label{fig:error}
\end{figure}

We add an example of error analysis to show in which case each segmentation method fails to segment the input file into an individual cough. Figure \ref{fig:error} depicts examples of a correct result and an error produced by each segmentation method. The correct and error results in each segmentation are annotated with a perfect agreement (5 out of 5). In this example, and maybe in most cases, all segmentation methods fail to segment the input file into an individual cough. The results of segmented coughs contain two coughs (for manual segmentation and hysteresis comparator methods) and three coughs (for the RMS threshold method). Future research could improve the sensitivity of individual cough segmentation regions against non-individual cough segmentation regions. The information in non-individual cough regions could be incorporated for building a binary classifier (individual cough vs. non-individual cough). In this study, the output of the evaluated segmentation methods is intended to contain only single-cough segmentation regions (it is not a binary classification but a segmentation task).

\section{Conclusions}
In this paper, we evaluated three different segmentation methods to extract individual (single) coughs from a given audio file containing several coughs. The evaluated methods are manual segmentation, hysteresis comparator, and RMS threshold. The manual segmentation method is performed by observing silence regions in the audio files, resulting in 121 files from the original 16 files. The automatic segmentation methods are performed by using a hysteresis comparator and RMS thresholds, resulting in 120 files and 151 files, respectively. We found the agreement of annotators to label the correct individual cough segmentations is fair for the manual segmentation and the hysteresis comparator and moderate for the RMS threshold method. The performance evaluation showed that the hysteresis comparator method gained better precision scores than the other two evaluated methods to segment into individual coughs. Finally, there is a large gap in precision scores between manual and automatic segmentation methods (with automatic is higher than manual), showing the potential research directions of automatic individual cough segmentation and cough counter. The future work will also include an evaluation of these segmentation methods for classifying cough-related diseases.

\section{Acknowledgements}
This paper is based on results obtained from a project, TM/DRPM-ITS/PN.02.003 (Skema Penelitian Keilmuan Sumber Dana ITS), Indonesia. The authors would like to thank students of the Engineering Physics Department ITS: Christina Mega Putri, Citra Annisaa Nurul Ain, Dhea Tirta Ananta, Muhammad Naufal Al Farros, Nindya Eka Winasis, and Raﬁ Surya Ghany, for annotating the segmented audio files. Parts of this research were supported by the New Energy and Industrial Technology Development Organization (NEDO), Japan, under Project No. JPNP20006.

\section*{Conflict of Interest}
On behalf of all authors, the corresponding author states that there is no conflict of interest.


\bibliography{cough}% common bib file


\end{document}